\RequirePackage{etoolbox}
\newtoggle{conference}

 \togglefalse{conference} % For the arXive version

\newcommand{\inConference}[1]{\iftoggle{conference}{#1}{}} % For text that should appear in the conference format only
\newcommand{\inArXiv}[1]{\iftoggle{conference}{}{#1}}  % For text that should appear in the Arxiv format only
\inArXiv{\documentclass[11pt]{article}}
\inConference{\documentclass[a4paper,UKenglish,numberwithinsect]{lipics-v2016}}
\inArXiv{
	\usepackage{graphicx}
	\usepackage{latexsym}
	\usepackage{amssymb}
	\usepackage{amsmath}
	\usepackage{amsthm}
}
\usepackage{forloop}
\inArXiv{\usepackage[paper=letterpaper,margin=1in]{geometry}}
\usepackage[ruled,vlined,linesnumbered]{algorithm2e}
\usepackage{paralist}
\usepackage{nicefrac}
\usepackage{caption}
\inArXiv{\usepackage{hyperref}}

\DeclareCaptionType{mechanism}[Mechanism][List of Mechanisms]
\inArXiv{
	\newtheorem{theorem}{Theorem}[section]
	\newtheorem{lemma}[theorem]{Lemma}
	\newtheorem{corollary}[theorem]{Corollary}
	
}
\inConference{\theoremstyle{plain}}
\newtheorem{observation}[theorem]{Observation}
\inArXiv
{
	\newcommand{\itemizeEnvironment}{compactitem}
	\newcommand{\enumerateEnvironment}{compactenum}
}
\inConference
{
	\newcommand{\itemizeEnvironment}{itemize}
	\newcommand{\enumerateEnvironment}{enumerate}
}

\makeatletter
\newtheorem*{rep@theorem}{\rep@title}
\newcommand{\newreptheorem}[2]{%
\newenvironment{rep#1}[1]{%
 \def\rep@title{#2 \ref{##1}}%
 \begin{rep@theorem}}%
 {\end{rep@theorem}}}
\makeatother

\newreptheorem{theorem}{Theorem}
\newreptheorem{observation}{Observation}
\newreptheorem{lemma}{Lemma}

\newcommand{\defcal}[1]{\expandafter\newcommand\csname c#1\endcsname{{\mathcal{#1}}}}
\newcommand{\defbb}[1]{\expandafter\newcommand\csname b#1\endcsname{{\mathbb{#1}}}}
\newcounter{calBbCounter}
\forLoop{1}{26}{calBbCounter}{
    \edef\letter{\Alph{calBbCounter}}
    \expandafter\defcal\letter
		\expandafter\defbb\letter
}

\newcommand{\GfT}{\mathtt{GfT}}
\newcommand{\Val}{\mathtt{Val}}
\newcommand{\OPM}{{\texttt{OPM}}}
\newcommand{\ie}{{\it i.e.}}
\newcommand{\eg}{{\it e.g.}}
\newcommand{\keywordsString}{Online dynamic mechanisms, online advertising market, mutli-sided markets, gain from trade maximization}
\makeatletter \newcommand{\hypertargettop}[1]{\Hy@raisedlink{\hypertarget{#1}{}}}\makeatother

\newcommand{\titleString}{Online Truthful Mechanisms for Multi-sided Markets}
\title{\inConference{\titleString}\inArXiv{\textbf{\titleString}}\footnote{This work is supported by the Horizon 2020 funded project TYPES (Project number: 653449. Call Identifier H2020-DS-2014-1). 
\inConference{We are submitting this paper for confidential review to be considered for publication in ESA on Aug 22--16, 2016.}\inArXiv{We are submitting this paper for confidential review to be considered for publication in 2017.}}}
\inConference{\titlerunning{\titleString}}

\inArXiv{
\author{
 Moran Feldman\thanks{Department of Mathematics and Computer Science, the Open University of Israel. E-mail: \href{mailto:moranfe@openu.ac.il}{moranfe@openu.ac.il}.}
 \and
 Rica Gonen\thanks{Department of Management and Economics, the Open University of Israel. E-mail: \href{mailto:gonenr@openu.ac.il}{gonenr@openu.ac.il}}
}
}

\inConference{
\author[1]{Moran Feldman}
\author[2]{Rica Gonen}
\affil[1]{Department of Mathematics and Computer Science, the Open University of Israel\\
  \texttt{moranfe@openu.ac.il}}
\affil[2]{Department of Management and Economics, the Open University of Israel\\
  \texttt{gonenr@openu.ac.il}}
\authorrunning{M. Feldman and R. Gonen}
	
\Copyright{Moran Feldman and Rica Gonen}

\keywords{\keywordsString}

\subjclass{F.1.2 Modes of Computation---Online computation, F.2.2 Nonnumerical Algorithms and Problems, J.4 Social and Behavioral Sciences---Economics, K.4.4 Electronic Commerce}
}

\begin{document}

\maketitle
%\documentclass[10pt]{article}
%\usepackage[usenames]{color} %used for font color
%\usepackage{amssymb} %maths
%\usepackage{amsmath} %maths
%\usepackage[utf8]{inputenc} %useful to type directly diacritic characters
%\begin{document}
\begin{abstract}

The study of mechanisms for multi-sided markets has received an increasingly growing attention from the research community, and is motivated by the numerous examples of such markets on the web and in electronic commerce, including: online advertising exchanges, stock exchanges, business-to-business commerce and bandwidth allocation. Many of these examples represent dynamic and uncertain environments, and thus, require, in fact, online mechanisms. Unfortunately, as far as we know, no previously published online mechanism for a multi-sided market (or even for a double-sided market) has managed to (approximately) maximize the gain from trade, while guaranteeing desirable economic properties such as incentivizing truthfulness, voluntary participation and avoiding budget deficit.

In this work we present the first online mechanism for a multi-sided market which has the above properties. Our mechanism is designed for a market setting suggested by [Feldman and Gonen (2016)]; which is motivated by the foreseeable future form of online advertising. To overcome privacy concerns, the model of [Feldman and Gonen (2016)] introduces offline user information markets through information brokers into the online advertising ecosystem, and thus, gives users control over which parts of their data get shared in the online advertising market. We note that both the advertisers and the information brokers are multi-minded players in this model.%We design an {\bf online} mechanism for [Feldman and Gonen (2016)]'s market model. Our online mechanism introduces a new pricing scheme concept where users get paid in ongoing increments up to a bound. The bound is pre-known yet the increments and total payment are dependent on the market economic demand and supply. Our newly suggested pricing scheme concept resembles known schemes used in practice such as executive compensation payments and provides independent interest.

The online nature of our setting motivated us to define a stronger notion of individual rationality, called ``continuous individual rationality'', capturing the natural requirement that a player should never lose either by participating in the mechanism \emph{or} by not leaving prematurely. Satisfying the requirements of continuous individual rationality, together with the other economic properties our mechanism guarantees, requires the mechanism to use a novel pricing scheme where users may be paid ongoing increments during the mechanism's execution up to a pre-known maximum value. As users rarely ever get paid in reality, this pricing scheme is new to mechanism design. Nevertheless, the principle it is based on can be observed in many common real life scenarios such as executive compensation payments and company acquisition deals. We believe both our new dynamic pricing scheme concept and our strengthened notion of individual rationality are of independent interest. 

%Our online mechanism introduces a new pricing scheme concept where users may be paid ongoing increments during the algorithm's execution and up to a bounded price. The bounded price is pre-known yet the increments and total payment are not pre-know and depend on the market's online changing demands and supplies. Though our pricing scheme is new to mechanism design and in reality users rarely ever get paid, the principle and economic logic of the scheme can be observed in real life scenarios such as executive compensation payments. Furthermore, since our newly introduced pricing results in a potential players' utility change over time, we extended and strengthen the property of individual rationality to include the possibility of such dynamic pricing. Both our new dynamic pricing scheme concept and our extended property of individual rationality provide independent interest. 

\medskip
\noindent \textbf{Keywords:} \keywordsString
\end{abstract}
%\end{document}

\inArXiv{
\pagenumbering{Alph}
\thispagestyle{empty}
\newpage
\setcounter{page}{1}
\pagenumbering{arabic}
}

\newcommand{\incentivesForUsersProof}
{
\begin{proof}
If $m$ arrives during the observation phase (\ie, $m \in M_T$), then no user of $m$ is ever assigned to a slot or paid. Hence, the lemma is trivial in this case. Thus, we assume in the rest of the proof that $m$ arrives after the observation phase.

Note that {\OPM} calculates the threshold $c(\hat{p})$ based on the reports of advertisers and mediators in $A_T$ and $M_T$, respectively. Thus, $p$, who is associated with a mediator not belonging to $M_T$, cannot affect this threshold. Next, let us denote by $k$ the number of users of $m$ that are assigned to slots when $p$ reports a cost smaller than $c(\hat{p})$. We claim that $k$ is independent of the exact cost reported by $p$, as long as this cost is smaller than $c(\hat{p})$. The reason for that is that most of the time {\OPM} accesses the reported cost of $p$ only by checking whether $p$ is assignable, and the answer for that check does not change as long as the reported cost of $p$ is smaller than $c(\hat{p})$. The exact value of $c(p)$ is only used by {\OPM} after {\OPM} decides to assign \emph{some} user of $m$ to a slot, and then this exact value is used to decide which user of $m$ will be assigned to the slot---which does not affect $k$.

Let $p'$ be the user of $m$ with the $k^{th}$ smallest cost among his users that are not $p$. If $m$ does not have $k$ users other than $p$, then $p'$ is a dummy user of cost $\infty$. In the rest of the proof we show that $p$ is assigned to a slot if and only if she reports a cost smaller than $\min\{c(\hat{p}), c(p')\}$. Moreover, when $p$ is assigned to a slot the total payment she gets is this minimum (which is her critical value). Clearly, the incentive compatability of {\OPM} for $p$ follows immediately from this claim.

Let us begin proving the above claim by showing that $p$ is left unassigned when she reports a cost larger than $\min\{c(\hat{p}), c(p')\}$. There are two cases to consider. If $p$ reports a cost larger than $c(\hat{p})$ then she is not assignable, and thus, she is left unassigned. On the other hand, consider the case that $p$ reports a value smaller than $c(\hat{p})$, but larger than $c(p')$. In this case $p$ is not one of the $k$ users of $m$ with the smallest reported costs, and thus, she is again left unassigned.

Next, we prove the other side of the above claim, \ie, that when $p$ reports a cost smaller than $\min\{c(\hat{p}), c(p')\}$ she is assigned and the total payment she gets is this minimum. The fact that $p$ reports a value smaller than $c(\hat{p})$ implies that $p$ is assignable, and the fact that she reports a value smaller than $c(p')$ guarantees that $p$ is one of the $k$ users of $m$ with the smallest reported costs. This already guarantees that $p$ is assigned to some slot, and that $p'$ is the unassigned user of $m$ with the smallest cost when {\OPM} updates for the last time the recommended total payment from $m$ to $p$ (unless $p'$ is a dummy user). Hence, the recommended total payment for $p$ is determined as follows. If $p'$ is assignable (\ie, $c(p') < c(\hat{p})$), then the recommended total payment for $p$ is set to $c(p')$. Otherwise, $m$ has no unassigned assignable users, and thus, the recommended total payment to $p$ is set to $c(\hat{p}) < c(p')$.

It remains to prove that {\OPM} is continuously individually rational for $p$, \ie, that the utility of $p$ can only increase over time when $p$ is truthful. The first time the utility of $p$ might change is when $p$ is assigned. When this happens $p$ is immediately payed an amount equal either to $c(\hat{p})$ or to the cost of an unassigned assignable user of $m$. Since {\OPM} chooses the user to assign as the unassigned assignable user of $m$ with the lowest cost, both possible payments are larger than $c(p)$, and thus, the utility of $p$ does not become negative following the assignment of $p$. Next, we prove that the recommended total payment for $p$ can only increase over time, which proves that $p$'s utility can only increase from the moment $p$ is assigned. To see why that is true, recall that, at every time point in which {\OPM} updates the recommended total payment to $p$, this total payment is updated to be either the cost of the unassigned assignable user of $m$ with the lowest cost, or $c(\hat{p})$ if $m$ has no unassigned assignable users left. As long as $m$ has unassigned assignable users this update rule yields a recommended total payment which can only increase over time since users occasionally get removed from the set of unassigned assignable users of $m$ (when they get assigned), but no user is ever added to this set. Moreover, the recommended total payment to $p$ also increases when the last unassigned assignable user of $m$ gets assigned since the recommended total payment to $p$ before this point was equal to the cost of some assignable user of $m$, which is smaller, by definition, than the new recommended total payment $c(\hat{p})$.
\end{proof}
} % End of: \incentivesForUsersProof

\newcommand{\incentivesForMediatorsProof}
{
\begin{proof}
If $m$ arrives during the observation phase (\ie, $m \in M_T$), then no user of $m$ is ever assigned to a slot and $m$ receives no payment. Hence, the lemma is trivial in this case. Thus, we assume in the rest of the proof that $m$ arrives after the observation phase.

Note that {\OPM} calculates the threshold $c(\hat{p})$ based on the reports of advertisers and mediators in $A_T$ and $M_T$, respectively. Thus, $m$, who does not belong to $M_T$, cannot affect this threshold. Whenever a user $p \in P(m)$ is assigned to a slot the utility of $m$ (and the user) decreases by $c(p)$ and increases by the additional payment $m$ gets, which is $c(\hat{p})$. In other words, the utility of $m$ changes by $c(\hat{p}) - c(p)$ (independently of the amount $m$ forwards to $p$). When $m$ is truthful this change is always non-negative since the fact that $p$ was assigned implies that she is assignable, \ie, her reported cost is smaller than $c(\hat{p})$. This already proves that each assignment of a user of $m$ increases his utility by a non-negative amount when he is truthful (assuming his users are also truthful), and thus, {\OPM} is continuously individually rational for $m$.

Let $s$ be the number of assignable users of $m$, according to his report. We claim that there exists a value $k$ which is independent of the report of $m$ such that for any report of $m$ the mechanism assigns the $\min\{k, s\}$ users of $m$ with the lowest reported costs. Before proving this claim, let us explain why the lemma follows from this claim. The above description shows that the utility of $m$ changes by a $c(\hat{p}) - c(p)$ for every assigned user $p \in P(m)$, thus, $m$ wishes to assign as many as possible users having cost less than $c(\hat{p})$, and if he cannot assign all of them then he prefers to assign the users with the lowest costs. By being truthful $m$ guarantees that only users of cost less than $c(\hat{p})$ are considered assignable, and thus, have a chance to be assigned. Moreover, by the above claim {\OPM} assigns the $k$ assignable users of $m$ with the lowest costs (or all of them if $s < k$), which is the best result $m$ can hope for given that at most $k$ of his users can be assigned. Hence, truthfulness is a dominant strategy for $m$.

We are only left to prove the above claim. Note that Observation~\ref{obs:invariant} implies that {\OPM} assigns no users of $m$ as long as there are mediators appearing earlier in $\sigma_E$ which still have unassigned assignable users. Once there are no more such mediators, {\OPM} assigns users of $m$, in an increasing costs order, to unassigned assignable slots till one of two things happens: either $m$ runs out of unassigned assignable users, or the input for {\OPM} ends. This means that when the input for {\OPM} ends before all the assignable users of mediators appearing before $m$ in $\sigma_E$ are assigned, then no users of $m$ are assigned and the claim holds with $k = 0$. Otherwise, we choose $k$ to be the number of unassigned assignable slots immediately before {\OPM} assigns the first user of $m$ (we count in $k$ both unassigned assignable slots of advertisers that have already arrived at this moment and unassigned assignable slots of advertisers that arrive later). Notice that the report of $m$ does not affect the behavior of {\OPM} up to the moment it starts assigning users of $m$, and thus, $k$ is independent of the report of $m$. If $s > k$, then the $k$ users of $m$ with the lowest costs are assigned before {\OPM} runs out of input and stops. Otherwise, if $s \leq k$ then {\OPM} stops assigning users of $m$ only after assigning all the assignable users of $m$, and thus, all the $s$ assignable users of $m$ get assigned.
\end{proof}
} % End of: \incentivesForMediatorsProof

\newcommand{\tildeElementsProof}
{
\begin{proof}
If $1 - 6r^{-1} \cdot \sqrt[3]{\alpha} \leq 0$, then both $\tilde{B}$ and $\tilde{P}$ are empty, and the inequality that we need to prove holds since its left hand side is $0$ and its right hand side is non-positive (recall that $S_c(P, B)$ is an assignment of users from $P$ to slots of $B$ maximizing the gain from trade, and thus, its gain from trade is at least $0$ since $\GfT(\varnothing) = 0$). Thus, we may assume in the rest of the proof that $1 - 6r^{-1} \cdot \sqrt[3]{\alpha} > 0$.

Since $\tilde{B}$ contains the $\lceil (1 - 6r^{-1} \cdot \sqrt[3]{\alpha})\tau \rceil$ slots with the largest values among the slots of $B_o$, we get:
\[
	\sum_{b \in \tilde{B}} v(b)
	\geq
	\lceil (1 - 6r^{-1} \cdot \sqrt[3]{\alpha})\tau \rceil \cdot \frac{\sum_{b \in B_o} v(b)}{\tau}
	\enspace.
\]
Similarly, since $\tilde{P}$ contains the $\lceil (1 - 6r^{-1} \cdot \sqrt[3]{\alpha})\tau \rceil$ users with the lowest costs among the users of $P_o$, we get:
\[
	\sum_{p \in \tilde{A}} c(p)
	\leq
	\lceil (1 - 6r^{-1} \cdot \sqrt[3]{\alpha})\tau \rceil \cdot \frac{\sum_{c \in P_o} c(p)}{\tau}
	\enspace.
\]
Combining the two inequities gives:
\begin{align*}
	\sum_{b \in \tilde{B}} v(b) - \sum_{p \in \tilde{P}} c(p)
	\geq{} &
	\lceil (1 - 6r^{-1} \cdot \sqrt[3]{\alpha})\tau \rceil \cdot \frac{\sum_{b \in B_o} v(b) - \sum_{p \in P_o} c(p)}{\tau}\\
	={} &
	\lceil (1 - 6r^{-1} \cdot \sqrt[3]{\alpha})\tau \rceil \cdot \frac{\GfT(S_c(P, B))}{\tau}\inConference{\\}
	\geq\inConference{{} &}
	(1 - 6r^{-1} \cdot \sqrt[3]{\alpha}) \cdot \GfT(S_c(P, B))
	\enspace.
	\qedhere
\end{align*}
\end{proof}
} % End of: \tildeElementsProof
%\documentclass[10pt]{article}
%\usepackage[usenames]{color} %used for font color
%\usepackage{amssymb} %maths
%\usepackage{amsmath} %maths
%\usepackage[utf8]{inputenc} %useful to type directly diacritic characters
%\begin{document}
\section{Introduction} \label{sec:introduction}

%Dynamic pricing mechanisms, and specifically
Mechanisms for multi-sided markets
% with multiple buyers and sellers
are given an increasingly growing focus in the research community. The study of such mechanisms is motivated by the continuous growth in the number of transactions and exchanges, and the need for competitiveness, which promotes adoption of new exchange mechanisms.
%dynamically run exchanges
Among the numerous examples of multi-sided markets on the web and in electronic commerce one can find: online advertising exchanges, stock exchanges, business-to-business commerce and bandwidth allocation. Many of these examples represent dynamic and uncertain environments, which led to an interest in online exchange mechanisms.

A natural expectation from an online exchange mechanism is to (approximately) maximize the gain from trade, while guaranteeing desirable economic properties such as incentivizing truthfulness, voluntary participation and avoiding budget deficit. Unfortunately, as far as we know, no previous work has managed to achieve these goals simultaneously. 
%Online versions of the double auction, where bids arrive (and in some models expire) at different times, were considered by \cite{BSZ02, WWW98,BPD07}.
Wurman et al.~\cite{WWW98} presented a mechanism incentivizing truthful reporting from either the buyers or the sellers, but not simultaneously from both. A different mechanism given by Blum et al.~\cite{BSZ02} maximizes the social welfare of buyers and non-selling sellers (as opposed to maximizing the gain from trade). Finally, Bredin et al.~\cite{BPD07} present a truthful online double-sided auction that is constructed from a truthful offline double-sided auction rule. However, the competitiveness of \cite{BPD07}'s mechanism with respect to the optimal trade was only studied empirically.

%\cite{BSZ02}'s algorithm assumes knowledge of an upper and lower bound on buyers'/sellers' valuations/costs

The failure of the above works to maximize the gain from trade while maintaining truthfulness, individual rationality (voluntary participation) and budget balance (avoiding budget deficit) can be partially attributed to an impossibility result of~\cite{MS83}. This impossibility result states that, even in an offline setting involving a single buyer and a single seller, maximizing the gain from trade while maintaining truthfulness and individual rationality perforce runs a deficit (\ie, is not budget balanced). An additional reason for the above failure is that the matching problem faced by the market maker (exchange mechanism) in multi-sided online markets combines elements of online algorithms and sequential decision making with considerations from mechanism design. More specifically, unlike in a traditional online algorithm, a mechanism for such a setting must provide incentives for players to report truthful information to the mechanism. On the other hand, unlike in traditional mechanism design, this is a dynamic setting with players that arrive over time, and the mechanism must deal with uncertainty and make irrevocable decisions before the arrival of all the players. 

In this work we present the first\footnote{To the best of our knowledge.} online mechanism for a multi-sided market setting which guarantees the economic properties of truthfulness, individual rationality, and budget balance while (approximately) maximizing the gain from trade. Specifically, we consider a market setting presented by~\cite{FG16}. 
This setting is motivated by online advertising in its foreseeable future form. Online advertising currently supports some of the most important Internet services, including: search, social media and user generated content sites. For online advertising to be effective, companies collect vast amounts of information about users, which increasingly creates privacy concerns. As these concerns are especially pronounced in the European society, EU regulators have actively been looking for ways to improve users' privacy. One such way, which was suggested by the EU regulators, is development of tools that enable the end user to choose which parts of their private information online advertising platforms are allowed to collect.

The market setting suggested by~\cite{FG16} based on this motivation includes advertisers as buyers, users as sellers (each willing to sell her own information portfolio through a broker) and information brokers as mediators representing the users. The objective of a mechanism for this setting is to end up with a match between users and advertisers maximizing the gain from trade. Towards that goal, the mechanism has to collect information from the mediators and advertisers; and thus, needs to incentivize the mediators and advertisers to report truthfully, which it can do by charging the advertisers and paying the mediators. Additionally, the mechanism can also recommend for the mediators to forward some of the payment they receive to the users, which allows the mechanism to affect the incentives of the users as well.% In addition to truthful reporting the mechanism's charges and payments may be designed such that no player can lose by participation, \ie, all players are individually rational. 

In order to convert the above offline setting suggested by~\cite{FG16} into an online setting, we assume the mediators and advertisers arrive at a uniformly random order, and refer to the arriving advertisers and mediators as arriving entities. Every time that a new entity arrives, the mechanism has an opportunity to assign users to advertisers. More specifically, when a mediator arrives the mechanism is allowed to assign users of the newly arriving mediator to advertisers that have already arrived. Similarly, when an advertiser arrives the mechanism is allowed to assign users of mediators that have already arrived to the newly arriving advertiser. Notice that this means that the mechanism is not allowed to cancel assignments that have already been made, or assign a user of a mediators that has already arrived to an advertiser that has already arrived. These restrictions, together with the random arrival order, represent the online nature of the setting. We note that our choice to model an online market using a random arrival order is a well established practice (see~\cite{BDGIT09,BIK07,VC16} for a few examples). Intuitively, this modeling choice is based on the assumption that real arrival orders are arbitrary rather than adversarial.

The online nature of our setting raises the question of what it means for a mechanism to be individually rational. As usual, individual rationality should imply that a player never losses by participating. However, in an online setting it is natural to require also that a player never losses by not leaving prematurely. We introduce a new concept called ``continuous individual rationality'' which captures the above intuitive requirement. Satisfying the requirements of continuous individual rationality, together with the other economic properties our mechanism guarantees, requires the mechanism to use a novel pricing scheme where users may be paid ongoing increments during the mechanism's execution. The maximum total payment that a user may end up with is pre-known (when the user arrives), however, the actual increments are not pre-know and depend on the market's online changing demands and supplies. As users rarely ever get paid in reality, this pricing scheme is new to mechanism design and might look odd at first glance. Nevertheless, the principle it is based on can be observed in many common real life scenarios such as executive compensation payments and company acquisition deals. For example, the eBay acquisition of Skype in 2005 involved both an upfront payment and an additional payment whose amount depended on the future performance of the bought company (\url{https://investors.ebayinc.com/releasedetail.cfm?releaseid=176402}).

%Incentivizing users to behave in a In order to convert the above online setting into a setting that can incentives users in a way that will maintain the desired economic properties, we introduce a new pricing scheme concept where users may be paid ongoing increments during the algorithm's execution and up to a bounded price. The bounded price is pre-known yet the increments and total payment are not pre-know and depend on the market's online changing demands and supplies. Though this pricing scheme is new to mechanism design and in reality users rarely ever get paid, the principle of the scheme can be observed in real life scenarios such as executive compensation payments. 

It is interesting to note that the setting of~\cite{FG16} involves multi-minded players (the advertisers and mediators), \ie, players with a multi-dimensional strategic space. The mechanisms suggested by~\cite{FG16} for this setting are the only known mechanisms, to date, which circumvent the impossibility result of \cite{MS83} in a model involving multi-minded players. Our result shows that it is possible to handle multi-minded players also in an online setting.

%To better understand the design challenge of our problem,  All the mechanisms circumventing \cite{MS83}'s impossibility (whether they are online or offline solutions) suggested in the literature to date\footnote{Excluding \cite{FG16} which will be discussed in the next subsections} only allow players with a single-dimensional strategic space, \ie, a single privately know parameter that the players can strategically act upon. In our model both advertisers and mediators have multi-dimensional strategic space as they can manipulate their reports of both their value/cost and their users' capacity/number respectively. 
%Another observed challenge of our problem is that the mediators control the information flow from their users to the mechanism. Therefore the mechanism faces strategic behavior not only from single players but also from mediators who represent essentially collusion sets of users: a mediator acts strategically to maximize utility on behalf of the users he represents.

\subsection{Our Result} \label{ssc:result}

%We design the first online multi-sided mechanisms able to handle players with multi-dimensional strategic spaces. Our mechanisms guarantee the desirable economic properties of truthfulness as well as individual rationality and budget balance, while yielding a gain from trade that approximates the optimal offline gain from trade. Furthermore, the  mechanism presented is the first mediated mechanism designed to run in an online fashion as well as the first multi-sided market with multi-dimensional strategic space to run online.

%As discussed above, our mechanism introduces a new dynamic pricing scheme where users' payments may be updated during the algorithm's execution stage up to a bounded price. Since our newly introduced pricing concept results in players' (the users) utility that may change over time, the property of individual rationality should be extended to include the possibility of such dynamic pricing. 

As discussed above, we extend the standard concept of individual rationality in a way that seems to match better the intuitive meaning of individual rationality in an online setting. Specifically, we say that a mechanism is \emph{continuously individually rational} for a player (a user, a mediator or an advertiser) if the player's utility can only increase over time when the player is truthful\footnote{Informally, a player is truthful if he/she reports the information as it is known to him/her. A formal definition of what does it mean for a user, mediator or advertiser to be truthful is given in Section~\ref{sec:model}.}. In other words, the utility of the player is always $0$ before the arrival of the first entity. After the arrival of each entity the mechanism updates the assignment and payments, which can affect the player's utility. If the change in the utility is always non-negative when the player is truthful, then the mechanism is continuously individually rational for the player. Note that the newly presented concept of continuous individual rationality is a stronger concept than individual rationality in its classic form as it requires two things. First, ex post individual rationality, \ie, the players are never worse off when the mechanism terminates, compared to their situation before the mechanism's execution. Second, it is individually rational for a player to remain \emph{throughout} the execution of the mechanism, \ie, each individual step during the execution can only increase the utilities of the players.

Like in \cite{BFT16, FG16}, we say that a mechanism is \emph{user-side incentive compatible} if truthfulness is a dominant strategy\footnote{Here and throughout the paper, a reference to domination of strategies should be understood as a reference to weak domination. We never refer to strong domination.} for each user given that her mediator is truthful (regardless of any parameter of the model, such as the number of mediators, and regardless of the other players' strategies). Similarly, the mechanism is \emph{user-side continuously individually rational} if it is continuously individually rational for each user given that her mediator is truthful (again regardless of any parameters and regardless of other players' strategies). A mechanism is \emph{mediator-side incentive compatible} if truthfulness is a dominant strategy for each mediator whose users are all truthful, and it is \emph{mediator-side continuously individually rational} if it is continuously individually rational for every such mediator (again regardless of any parameters or other players' strategies). Finally, a mechanism is \emph{advertiser-side incentive compatible} if truthfulness is a dominant strategy for every advertiser, and it is \emph{advertiser-side continuously individually rational} if it is continuously individually rational for every advertiser. We construct a mechanism which is \emph{three-sided incentive compatible} (\ie, it is simultaneously user-side incentive compatible, mediator-side incentive compatible and advertiser-side incentive compatible) and also \emph{three-sided continuously individually rational} (\ie, it is simultaneously user-side continuously individually rational, mediator-side continuously individually rational and advertiser-side continuously individually rational). 

Our mechanism is termed ``Observe and Price Mechanism'' ({\OPM}).
% and it works as follows. First, the mechanism observes a random subset of the arriving entities. The optimal trade for the observed set is then used to produce a threshold cost and a threshold value that allow budget balanced pricing. Build a stack of entities for every arriving entity after the observation phase and add every arriving entity that is not yet assigned.
%As a mediator's users can be assigned at different stages of the algorithm's execution with the arrival of new advertisers, the mechanism's recommendation for the mediators' payments to their users is constantly updated. 
The following theorem analyzes the economic properties guaranteed by {\OPM} and its competitive ratio. %shows that the above online mechanism is three-sided truthful, continuously individually rational, budget balanced and provides a non-trivial approximation for the offline optimal gain from trade.
The parameter $\alpha$ is an upper bound, known to the mechanism, on the market importance of any single player. Formally, $\alpha$ bounds the ratio between the size of the optimal trade and the maximum capacity of an advertiser or the maximum number of users that a mediator can represent. For large markets, such as the market motivating our work, $\alpha$ is expected to be much smaller than $1$.
  
\begin{theorem} \label{thm:OPM}
{\OPM} is budget balanced\footnote{A mechanism is \emph{budget balanced} if the amount it charges (from the advertisers) is at least as large as the amount it pays (to the mediators).}, three-sided continuously individually rational, three-sided incentive compatible and $(1 - 9.5\sqrt[6]{\alpha} - 10e^{-2/\sqrt[3]{\alpha}})$-competitive.
\end{theorem}

Note that our mechanism is randomized, but is guaranteed, by its three-sided incentive compatibility, to be truthful for all possible random coin flips. This stands in contrast to many randomized mechanisms from the literature, which are typically proven to be truthful only in expectation or with high probability.

\subsection{Additional Related Work}

Our setting is based on the model of~\cite{FG16}. From a motivational point of view this model is closely related to works that involve mediators and online advertising markets, such as~\cite{AMT09, FMMP10, SGP14}. The models studied by these works are motivated by the current networks exchanges for online display ads, where advertisers buy through mediators (the networks) which in turn buy goods (publishing space) from a single seller. Despite their motivation by a network exchange, these models are actually auctions (\ie, one-sided mechanisms). Moreover, they focus on offline mechanisms maximizing revenue, which is very different from the focus of our work. % they are offline auctions and therefore need not deal with the challenges and impossibility integrated by the multi-sided structure of our market, the requirement to keep the mechanism from running into a deficit and the challenges formed by the online uncertainty. Our focus is maximization of the gain from trade, unlike \cite{AMT09,FMMP10,SGP14} which focus on revenue maximization. %The closest model to our mediated model is \cite{FG16}. They too analyze an advertising mediated market where users sell information through mediators. They present two mechanisms that maintain all the properties we desire; however, both of their solutions are offline mechanisms that require all the biding information up front and leave the realm of uncertainty outside their model. An online solution as given in our current work fits better with the nature of the online adverting market problem.  
See~\cite{FG16} for additional works related to their model, and in particular to the study of mediators.

The uniformly random arrival order of the entities in our setting relates it to the vast literature on the secretary problem. The original form of the secretary problem first appeared around the 1960's, although its exact origin remains unclear~\cite{D63, F89, L61}. In this form, the problem asks to select online the maximum value element among a set of elements arriving at a uniformly random order. Connections found between various extensions of the secretary problem and mechanism design have motivated an extensive study of these extensions over the last decade (for an excellent survey on these extensions see~\cite{D13}). One extension of the secretary problem which is particularly relevant to our work is an extension in which the arriving elements are the right side nodes of a bipartite graph whose left side is known to the algorithm. The algorithm then have to construct a maximum weight matching online. More specifically, for every arriving right side node the algorithm must decide, immediately and irrevocably, whether to assign it to a left side node, and if so to which one. Kesselheim et al.~\cite{KRTV2013} describe a $e^{-1}$-competitive algorithm for this extension of the secretary problem. Our own setting can be viewed as a variant of this extension involving strategic considerations.

Our presented mechanism is composed of two phases: an observation phase in which the algorithm collects information but makes no assignments, and a matching phase. This partition of the mechanism into two phases is similar to the structure found in many algorithms for the secretary problem and its extensions. Moreover, within the literature on algorithmic game theory, an analog for the observation phase can be found in the work of Goldberg et al.~\cite{GHW01} who described random sampling auctions in which an observed (sampled) set is used in order to compute a threshold for the remaining players. It is important to note, however, that, unlike in our work, \cite{GHW01} focuses on one-sided online auctions with unlimited supply.

In addition to the above described extension of the secretary problem, there has been a significant body of works studying online matching problems with an adversarial arrival order. This body of work was originated by the work of Karp et al.~\cite{KVV90} who described an optimal online algorithm for unweighted bipartite online matching. Later works considered more general settings allowing various kinds of weights (see, for example,~\cite{AGKM11,CHN14,GM07}). We note that these works do not refer to strategic considerations despite the fact that they are mostly motivated by online advertisement markets.

Last but not least, we would like to mention matching markets, which are a model related to double-sided auctions---though without money. There is extensive literature on online matching markets. The classic matching algorithm is the deferred acceptance algorithm \cite{GS62}. This algorithm is truthful for one side of the market, and produces a stable match with respect to reported preferences. Moreover, it is known that there does not exist a stable matching mechanism that is truthful for all players \cite{R82}. Readers are referred to \cite{SU08} for more background on the matching markets literature.

%In general, our work relates to the study of mediators. Typically the work on mediators refers to a model where every mediator serves as an arbitration device: the players are not a captive audience, and each of them can decide to participate in the game directly or work through the mediator \cite{AMT09, MT09}. Similar to \cite{BFT16}, in our setting there are multiple mediators, each having his own captive audience which must play the game through the mediator. Moreover, the mediators are strategic players who are interested in maximizing their utility. As our mediators are not revenue maximizing but rather utility maximizing one can view them as collusion sets. In this context, a relevant work is \cite{LST02} which studies game theoretic aspects of bidding clubs in which "collusion devices`` (cartels) are strategically created in a first price auction. Unlike \cite{LST02}'s model, in our setting the partition of users to mediators is pre-determined and our focus is on mechanism design given that fact.

%\end{document}
\section{Model and Definitions} \label{sec:model}

Let us now present the exact details of the model we consider. The model consists of a set $P$ of users, a set $M$ of mediators, and a set $A$ of advertisers. Each user $p \in P$ has a non-negative cost $c(p)$ which she suffers if she is assigned to an advertiser; thus, the utility of $p$ is $0$ if she is not assigned and $t - c(p)$ if she is assigned and paid $t$. The users are partitioned among the mediators, and we denote by $P(m) \subseteq P$ the set of users associated with mediator $m \in M$ (\ie, the sets $\{P(m) \mid m \in M\}$ form a disjoint partition of $P$). The utility of a mediator $m \in M$ is the amount he is paid minus the total cost his users suffer; hence, if $x(p) \in \{0, 1\}$ is an indicator for the event that user $p \in P(m)$ is assigned and $t$ is the payment received by $m$ (part of which might have been forwarded by the mediator to his users), then the utility of $m$ is $t - \sum_{p \in P(m)} x(p) \cdot c(p)$. Finally, each advertiser $a \in A$ has a positive capacity $u(a)$, and she gains a non-negative value $v(a)$ from every one of the first $u(a)$ users assigned to her; thus, if advertiser $a$ is assigned $n \leq u(a)$ users and has to pay $t$ then her utility is $n \cdot v(a) - t$.

As explained in Section~\ref{sec:introduction}, we assume the entities (\ie, the mediators and advertisers) arrive at a uniformly random order. A mechanism for this model knows the total number of entities\footnote{In some cases the assumption that the mechanism has a prior knowledge about the number of entities might be considered unnatural. The mechanism we present can be modified using standard techniques to work with an alternative assumption stating that each entity arrives at a uniformly random time from some range (for example, $[0, 1]$). We refer the reader to~\cite{FNS11} for more details about the relation between these assumptions.}, and views the entities as they arrive; however, it has no prior knowledge about the parameters of the entities or about the users. To compensate for this lack of knowledge, each arriving entity reports information to the mechanism. Each advertiser reports her capacity and value. The reports of the mediators are formed in a slightly more involved way. Each user reports her cost to her mediator, and based on these reports each mediator reports the number of his users and their costs to the mechanism. The users, mediators and advertisers are all strategic, and thus, free to produce incorrect reports. In other words, an advertiser may report incorrect capacity and value, a user may report an incorrect cost and a mediator may report any number of users and associate with each one of them an arbitrary cost.

Every time that a new entity arrives, the mechanism has an opportunity to assign users to advertisers. More specifically, when a mediator arrives the mechanism is allowed to assign users of the newly arriving mediator to advertisers that have already arrived. Similarly, when an advertiser arrives the mechanism is allowed to assign users of mediators that have already arrived to the newly arriving advertiser. The objective of the mechanism is to end up with an assignment of users to advertisers maximizing the \emph{gain from trade}. In order to incentivize the mediators and advertisers to report truthfully, the mechanism may charge the advertisers and pay the mediators. Additionally, the mechanism is also allowed to recommend for each mediator how much of the payment he received to forward to each one of his user. It is important to observe that the utility function of the mediators is not affected by the forwarding of payments to the users, and thus, it is reasonable to believe that mediators follow the forwarding recommendations.

We say that a user is \emph{truthful} if she reports her true cost. Similarly, an advertiser is \emph{truthful} if she reports her true capacity and value. Finally, a mediator is considered \emph{truthful} if he reports to the mechanism his true number of users and the costs of the users as reported to him; and, in addition, he also pays the users according to the recommendation of the mechanism (in other words, he lets them know about their true balance).
%A mediator \emph{follows the mechanism} if his strategy implies the following:
%\begin{compactitem}
	%\item He reports his true number of users.
	%\item For every user he reports a cost equal to the cost reported by this user (\ie, he simply forwards the information he received from the users).
	%\item He forwards payments to the users as requested by the mechanism.
%\end{compactitem}
%A mechanism is \emph{continuously IR} for a player (a user, a mediator or an advertiser) if the utility of the player can only increase over time when the player is truthful (or follows the protocol when the player is a mediator)******Rica: we should define the truthfulness of a mediator as in the deterministic paper, i.e., truthful mediator is the one who reports costs and amount truthfully and follow the protocol payments to his users. Meaning the () are not needed.***********. In other words, the utility of the player is always $0$ before the arrival of the first entity. After the arrival of each entity the mechanism updates its assignment, which can affect the utility of the player. If the change in the utility is always non-negative when the player is truthful (or follows the protocol), then the mechanism is continuously IR for the player.% More generally, a mechanism is continuously IR if it is continuously IR for all the agents. Note that a continuously IR mechanism is in particular IR.

Similarly to the practice of~\cite{FG16}, we associate a set $B(a)$ of $u(a)$ slots with each advertiser $a \in A$. This allows us to think of the users as assigned to slots instead of directly to advertisers. Formally, let $B$ be the set of all slots (\ie, $B = \bigcup_{a \in A} B(a)$), then an assignment is a set $S \subseteq B \times P$ in which no user or slot appears in more than one ordered pair. We say that an assignment $S$ assigns a user $p$ to slot $b$ if $(p, b) \in S$. Similarly, we say that an assignment $S$ assigns user $p$ to advertiser $a$ if there exists a slot $b \in B(a)$ such that $(p, b) \in S$. It is also useful to define values for the slots. For every slot $b$ of advertiser $a$, we define the value $v(b)$ of $b$ as equal to the value $v(a)$ of $a$. Using this notation, the gain from trade of assignment $S$ can be stated as:
\[
	\GfT(S)
	=
	\sum_{(p, b) \in S} [v(b) - c(p)]
	\enspace.
\]

Finally, we would like to define two additional shorthands that we use occasionally. Given a set $A' \subseteq A$ of advertisers, we denote by $B(A') = \bigcup_{a \in A'} B(a)$ the set of slots belonging to advertisers of $A'$. Similarly, given a set $M' \subseteq M$ of mediators, $P(M') = \bigcup_{m \in M'} P(m)$ is the set of users associated with mediators of $M'$.

\subsection{Comparison of Costs and Values} \label{sec:comparison}

The presentation of our mechanism is simpler when the values of slots and the costs of users are all unique. Clearly, this is extremely unrealistic since all the slots of a given advertiser have the exact same value in our model. Thus, we simulate uniqueness using a tie-breaking rule. The rule we assume works as follows:
\begin{compactitem}
	\item The mechanism chooses an arbitrary order on the mediators and advertisers. It is important that this order is chosen independently of the reports received by the mechanism and the arrival order of the entities (if there is no natural order that can be used, the mechanism can simply choose a uniformly random order). The mechanism then uses this order to break ties when comparing users to slots and when comparing between users (slots) associated with different mediators (advertisers). For example, when comparing the cost of user $p$ with the value of a slot $b$, the mechanism breaks ties in favor of $p$ if and only if the mediator of $p$ appears earlier than the advertiser of $b$ in the chosen order.
	%\begin{compactitem}
		%\item When comparing the cost of user $p$ with the value of a slot $b$, the mechanism breaks ties in favor of $p$ if and only if the mediator of $p$ appears earlier than the advertiser of $b$ in $\sigma$.
		%\item When comparing the costs of two users $p_1$ and $p_2$ associated with \emph{different} mediators, the mechanism breaks ties in favor of $p_1$ if and only if the mediator of $p_1$ appears earlier than the mediator of $p_2$ in $\sigma$.
		%\item When comparing the values of two slots $b_1$ and $b_2$ associated with \emph{different} advertisers, the mechanism breaks ties in favor of $b_1$ if and only if the advertiser of $b_1$ appears earlier than the advertiser of $b_2$ in $\sigma$.
	%\end{compactitem}
	\item We assume that the report of every mediator induces some order on the set of users of this mediators. The mechanism uses this order to break ties between the costs of users belonging to the same mediator.% More specifically, when comparing two users $p_1$ and $p_2$ of the same mediator, the mechanism breaks the tie according  to the order induced by the report of this mediator.
	\item Finally, since the slots of a given advertiser are all identical and non-strategic (recall that slots were introduced into the model just for the purpose of simplifying the presentation), any method can be used for tie-breaking between the slots of a given advertiser.%The mechanism chooses an arbitrary order for the slots of each advertiser. Since the slots of a given advertiser are all identical, it is not important how this order is selected. When comparing values of two slots belonging to the same advertiser (which always have the same value), the mechanism breaks the tie according to the above order.
\end{compactitem}

In the rest of this paper, whenever costs/values are compared
%unless it is explicitly specified that they are compared as numbers,
the comparison is assumed to use the above tie breaking rule. Note that this assumption implies that two values (costs) are equal if and only if they belong to the same slot (user); moreover, the value of a slot is never equal to the cost of a user.
%We now prove a useful observation that follows from the way we defined the tie-breaking rule.
%
%\begin{observation}
%When the slots are ordered in an increasing (or decreasing) value order, all the slots of a single advertiser are always consecutive.
%\end{observation}
%\begin{proof}
%Let $b_1$ and $b_2$ be two slots of one advertiser, and let $b$ be an arbitrary slot of another advertiser. It is enough to prove that $v(b_1) < v(b)$ implies $v(b_2) < v(b)$. The inequality $v(b_1) < v(b)$ can happen in two cases. If $v(b_1)$ is smaller than $v(b)$ as numbers, then we get $v(b_2) < v(b)$ since $v(b_1)$ and $v(b_2)$ are equal as numbers. Otherwise, if $v(b_1)$ is equal to $v(b)$ as numbers then the inequality $v(b_1) < v(b)$ implies that the advertiser of $b_1$ (and $b_2$) appears earlier in $\sigma$ than the advertiser of $b$, and thus, we also have $v(b_2) < v(b)$.
%\end{proof}

\subsection{Canonical Assignment}

Given a set $B' \subseteq B$ of users and a set $P' \subseteq P$ of slots, the canonical assignment $S_c(P', B')$ is the assignment constructed by the following process. First, we order the slots of $B'$ in a decreasing value order $b_1, b_2, \dotsc, b_{|B'|}$ and the users of $P'$ in an increasing cost order $p_1, p_2, \dotsc, p_{|P'|}$. Then, for every $1 \leq i \leq \min\{|B'|, |P'|\}$ the canonical assignment $S_c(B', P')$ assigns user $p_i$ to slot $b_i$ if and only if $v(b_i) > c(p_i)$.

The canonical assignment is an important tool we use often in the next section. In some places we refer to the user or slot at location $i$ of a canonical assignment $S_c(P', B')$, by which we mean user $p_i$ or slot $b_i$, respectively. Additionally, the size $|S_c(P, B)|$ of the canonical assignment $S_c(P, B)$ is used very often in our proofs, and thus, it is useful to define the shorthand $\tau = |S_c(P, B)|$.

The following lemma, which was proved by~\cite{FG16}, shows that the canonical assignment is always an optimal assignment.

\begin{lemma}[Lemma 2.2 of~\cite{FG16}]
The canonical assignment $S_c(P', B')$ maximizes\inConference{ the expression} $\GfT(S_c(P', B'))$ among all the possible assignments of users of $P'$ to slots of $B'$.
\end{lemma}
\section{Our Mechanism} \label{sec:machanism}

In this section we describe our online mechanism ``Observe and Price Mechanism'' (\OPM). {\OPM} assumes $|S_c(P, B)| > 0$, and that there exists a value $\alpha \in [|S_c(P, B)|^{-1}, 1]$, known to the mechanism, such that we are guaranteed that:
\[
	\frac{u(a)}{|S_c(P, B)|} \leq	\alpha \quad \forall\; a \in A
	\qquad \text{and}	\qquad
	\frac{|P(m)|}{|S_c(P, B)|} \leq	\alpha \quad \forall\; m \in M
	\enspace.
\]
In other words, $\alpha$ is an upper bound on how large can the capacity of an advertiser or the number of users of a mediator be compared to the size of the optimal assignment $S_c(P, B)$. We remind the reader that $\alpha$ can be informally understood as a bound on the market importance of every single advertiser or mediator. %For large markets, such as the market motivating our work, $\alpha$ is expected to be much smaller than $1$.

A description of {\OPM} is given as Mechanism~\ref{mch:OPM}. Notice that Mechanism~\ref{mch:OPM} accepts a parameter $r \in (0, \nicefrac{1}{2}]$ whose value is specified later. Additionally, Mechanism~\ref{mch:OPM} often refers to parameters of the model that are not known to the mechanism, such as the value of an advertiser or the number of users of a mediator. Whenever this happens, this should be understood as referring to the reported values of these parameters.

\vspace{2mm}
\noindent \begin{minipage}{\textwidth}
\captionsetup{type=mechanism}
\noindent \rule{\linewidth}{0.8pt}
\vspace{-6.5mm}\captionof{mechanism}{Observe and Price Mechanism (\OPM)}\label{mch:OPM}
\noindent \rule{\linewidth}{0.8pt}
\vspace{-6mm}
\begin{compactenum}[\bfseries 1.]
	\item Draw a random value $t$ from the binomial distribution $\cB(|A| + |M|, r)$, and observe the first $t$ entities that arrive without assigning any users. Let $A_T$ and $M_T$ be the set of the observed advertisers and mediators, respectively. More formally, if $T$ is the set of the first $t$ entities that arrived, then $A_T = A \cap T$ and $M_T = M \cap T$. We later refer to this step of the mechanism as the ``observation phase''.
\end{compactenum}
\end{minipage}
{
\hphantom{1} %\inArXiv{\vspace{-1.8mm}}
\setlength{\plitemsep}{3pt}
\begin{compactenum}[\bfseries 1.]
	\setcounter{enumi}{1}
	\item Let $\hat{p}$ and $\hat{b}$ be the user and slot at location $\lceil (1 - 2r^{-1} \cdot \sqrt[3]{\alpha}) \cdot |S_c(P(M_T), B(A_T))| \rceil$ of the canonical assignment $S_c(P(M_T), B(A_T))$. If $(1 - 2r^{-1} \cdot \sqrt[3]{\alpha}) \cdot |S_c(P(M_T), B(A_T))| \leq 0$, then the previous definition of $\hat{p}$ and $\hat{b}$ cannot be used. Instead define $\hat{p}$ as a dummy user of cost $-\infty$ and $\hat{b}$ as a dummy slot of value $\infty$. We say that a slot $b$ or a user $p$ corresponding to an entity that arrived \emph{after} the observation phase is \emph{assignable} if $v(b) > v(\hat{b})$ or $c(p) < c(\hat{p})$, respectively.
	
	\item Let $\sigma_E$ be the sequence of the entities that arrived so far after the observation phase. Initially $\sigma_E$ is empty, and entities are added to it as they arrive.
	
	\item For every arriving entity:
	\begin{compactenum}[\bfseries a.]
		\item Add the new entity to the end of $\sigma_E$.

		\item If the arriving entity is a mediator $m$, then, as long as $m$ has unassigned assignable users and there is an advertiser in $\sigma_E$ having unassigned assignable slots, do the following:
		\begin{compactitem}[$\bullet$]
			\item Let $a$ be the earliest advertiser in $\sigma_E$ having unassigned assignable slots.
			\item Assign the unassigned assignable user of $m$ with the lowest cost to an arbitrary unassigned assignable slot of $a$, charge an amount of $v(\hat{b})$ from advertiser $a$ and pay $c(\hat{p})$ to mediator $m$.
		\end{compactitem}

		\item If the arriving entity is an advertiser $a$, then, as long as $a$ has unassigned assignable slots and there is a mediator in $\sigma_E$ having unassigned assignable users, do the following:
		\begin{compactitem}[$\bullet$]
			\item Let $m$ be the earliest mediator in $\sigma_E$ having unassigned assignable users.
			\item Assign the unassigned assignable user of $m$ with the lowest cost to an arbitrary assignable slot of $a$, charge an amount of $v(\hat{b})$ from advertiser $a$ and pay $c(\hat{p})$ to mediator $m$.
		\end{compactitem}
		\item For every mediator $m \in \sigma_E$, recommend $m$ to transfer his assigned users an additional amount that guarantees the following:
		\begin{compactitem}[$\bullet$]
			\item If all the assignable users of $m$ are assigned to slots, then the additional amount should increase the total payment received so far by each assigned user of $m$ to $c(\hat{p})$.
		\end{compactitem}
	\end{compactenum}
\end{compactenum}
}
\inArXiv{\vspace{0.1cm}}\inConference{\vspace{0.12cm}}
\noindent\begin{minipage}{\textwidth}
\begin{compactenum}[\bfseries 1.]
	\item[\hphantom{\bfseries 1.}]
	\begin{compactenum}[\bfseries a.]
		\item[\hphantom{\bfseries a.}]
		\begin{compactitem}[$\bullet$]
			\item Otherwise, let $p$ be the unassigned assignable user of $m$ with the minimum cost. In this case the additional amount should increase the total payment received so far by each assigned user of $m$ to $c(p)$.\footnotemark
		\end{compactitem}
	\end{compactenum}
\end{compactenum}
\vspace{-3mm}\noindent \rule{\linewidth}{0.8pt}
\end{minipage}
\footnotetext{Note that at every point in time $m$ is budget balanced since he receives a payment of $c(\hat{p})$ for each one of his assigned users, and the total amount recommended for him to pay to each one of these users is either $c(\hat{p})$ or equal to the cost of some assignable user (and thus, is upper bounded by $c(\hat{p})$).}
\vspace{2mm}

We would like to note that {\OPM} is based on a mechanism of~\cite{FG16} named ``Threshold by Partition Mechanism'', and the analyses of both mechanisms go along similar lines. However, {\OPM} introduces additional ideas that allow it to work in an online setting. In particular, {\OPM} uses an involved recommended payments updating rule that keeps it three-sided continuously individually rational. Moreover, {\OPM} is able to use an observation phase whose size is a small fraction of the entire input (for $\alpha \ll 1$), whereas the analysis of the original mechanism of~\cite{FG16} relies on the symmetry properties induced by an even partition of the input (which is inappropriate in an online setting).

Let us start the analysis of {\OPM} with the following simple observation, which shows that {\OPM} obeys the restriction of our model on the way a mechanism may update its assignment.

\begin{observation}
Each time {\OPM} assigns a user to a slot, either the user belongs to the newly arrived mediator or the slot belongs to the newly arrived advertiser.
\end{observation}

Our objective in the rest of this section is to prove Theorem~\ref{thm:OPM}. In fact, we prove the following restatement of the theorem, which implies the original statement of the theorem from Section~\ref{ssc:result}.

\begin{reptheorem}{thm:OPM}
{\OPM} is budget balanced, three-sided continuously individually rational, three-sided incentive compatible and $(1 - r - 22r^{-1} \cdot \sqrt[3]{\alpha} - 10e^{-2/\sqrt[3]{\alpha}})$-competitive. Hence, for $r = \min\{1/2, 4\sqrt[6]{\alpha}\}$ the competitive ratio of {\OPM} is at least: $1 - 9.5\sqrt[6]{\alpha} - 10e^{-2/\sqrt[3]{\alpha}}$.
\end{reptheorem}

One part of Theorem~\ref{thm:OPM} (\ie, that {\OPM} is budget balanced) is proved by the following observation.

\begin{observation}
{\OPM} is budget balanced.
\end{observation}
\begin{proof}
We show that whenever {\OPM} assigns a user $p$ to a slot $b$, it charges the advertiser of $b$ more than it pays the mediator of $p$. Consider an arbitrary ordered pair $(p, b)$ from the assignment produced by {\OPM}. The fact that $p$ is assigned implies that $c(p) < c(\hat{p})$, and thus, $\hat{p}$ is not a dummy user (since $c(\hat{p}) = -\infty$ when $\hat{p}$ is a dummy user). Similarly, the fact that a user is assigned to $b$ implies that $v(b) > v(\hat{b})$, and thus, $\hat{b}$ is not a dummy slot (since $v(\hat{b}) = \infty$ when $\hat{b}$ is a dummy slot).

Recall that the fact that $\hat{p}$ and $\hat{b}$ are not dummy user and slot, respectively, implies that $\hat{p}$ and $\hat{b}$ are matched by the canonical assignment $S_c(P(M_T), B(A_T))$. Since a canonical assignment never assigns a user $p'$ to a slot $b'$ when $c(p') > v(b')$, we get: $c(\hat{p}) < v(\hat{b})$. The proof now completes by observing that the advertiser of $b$ is charged $v(\hat{b})$ for the assignment of $p$ to $b$, and the mediator of $p$ is paid only $c(\hat{p})$ for the assignment of $p$ to $b$.
\end{proof}

Following is a useful observation about {\OPM} that we occasionally use in the next proofs.

\begin{observation} \label{obs:invariant}
{\OPM} preserves the invariant that one of the following is always true immediately after {\OPM} processes the arrival of an entity:
\begin{\enumerateEnvironment}
	\item {\OPM} assigned all the assignable users of mediators that have already arrived. \label{case:1}
	\item {\OPM} assigned users to all the assignable slots of advertisers that have already arrived. \label{case:2}
\end{\enumerateEnvironment}
\end{observation}
\begin{proof}
Clearly the invariant holds during the observation phase because only mediators and advertisers that arrive after the observation phase contribute assignable users and slots, respectively. Next, assume the invariant held before the arrival of some mediator $m$ which arrives after the observation phase, and let us prove that it holds also after the arrival of $m$. If before the arrival of $m$ case~\eqref{case:2} of the invariant held, then this case also holds after the arrival of $m$ since $m$ contributes no new slots. On the other hand, if case~\eqref{case:1} held before the arrival of $m$, then {\OPM} assigns the assignable users of $m$ to assignable slots of advertisers that have already arrived till one of two things happen: either all the assignable slots of advertisers that have already arrived get assigned (and thus, case~\eqref{case:2} of the invariant now holds), or all the assignable users of $m$ get assigned (and thus, case~\eqref{case:1} of the invariant holds again). It remains to prove that if the invariant held before the arrival of an advertiser $a$ which arrives after the observation phase, then it also holds after her arrival. However, this proof is analogous to the above proof for mediators, and thus, we omit it.
\end{proof}

\subsection{The Incentive Properties of {\OPM}}

In this section we prove the incentive parts of Theorem~\ref{thm:OPM}. Specifically, we \inArXiv{prove}\inConference{give} three lemmata showing that {\OPM} is three-sided continuously individually rational and three-sided incentive compatible. The first lemma analyzes the incentive properties of {\OPM} for users.\inConference{ Due to space constraints, the proofs of this lemma and the next one have been deferred to Appendix~\ref{app:missing_proofs}.}

\begin{lemma} \label{lem:incentives_for_users}
For every user $p$, assuming the mediator $m$ of $p$ is truthful, {\OPM} is continuously individually rational for $p$, and truthfulness is a dominant strategy for her.
\end{lemma}
\inArXiv{\incentivesForUsersProof}

The next lemma analyzes the incentive properties of {\OPM} for mediators.

\begin{lemma} \label{lem:incentives_for_mediators}
For every mediator $m$, assuming the users of $m$ are truthful, {\OPM} is continuously individually rational for $m$, and truthfulness is a dominant strategy for him.
\end{lemma}
\inArXiv{\incentivesForMediatorsProof}

Finally, the next lemma considers the incentive properties of {\OPM} for advertisers. The proof of this lemma is analogous to the proof of the previous lemma (with slots exchanging roles with users, $v(\hat{b})$ exchanging roles with $c(\hat{p})$, etc.), and thus, we omit it.
\begin{lemma}
For every advertiser $a$, {\OPM} is continuously individually rational for $a$, and truthfulness is a dominant strategy for her.
\end{lemma}

\subsection{The Competitive Ratio of {\OPM}}

In this section we analyze the competitive ratio of {\OPM}. Recall that $\tau$ was defined as a shorthand for $|S_c(P, B)|$. We now define $\tilde{P}$ ($\tilde{B}$) as the set of the users (slots) at locations $1$ to $\lceil (1 - 6r^{-1} \cdot \sqrt[3]{\alpha})\tau \rceil$ of the canonical assignment $S_c(P, B)$ ($\tilde{P}$ and $\tilde{B}$ are defined to be empty when $1 - 6r^{-1} \cdot \sqrt[3]{\alpha} \leq 0$). The following observation shows that most of the gain from trade of the canonical assignment $S_c(P, B)$ comes from the users and slots of $\tilde{P}$ and $\tilde{B}$, respectively. For convenience, let us denote by $P_o$ the set of users that are assigned by $S_c(P, B)$, and by $B_o$ the set of slots that are assigned some user by $S_c(P, B)$.\inConference{ Due to space constraints, the proof of the next lemma has been deferred to Appendix~\ref{app:missing_proofs}.}

\begin{observation} \label{obs:tilde_elements}
$\sum_{b \in \tilde{B}} v(b) - \sum_{p \in \tilde{P}} c(p) \geq (1 - 6r^{-1} \cdot \sqrt[3]{\alpha}) \cdot \GfT(S_c(P, B))$.
\end{observation}
\inArXiv{\tildeElementsProof}

Observation~\ref{obs:tilde_elements} shows that one can prove a competitive ratio for {\OPM} by relating the gain from trade of the assignment it produces to the gain from trade obtained by assigning the users of $\tilde{P}$ to the slots $\tilde{B}$. The following lemma is a key lemma we use to relate the two gains. In order to state this lemma we need some additional definitions. Consider the following two sets.
\[
	\hat{P} = \{p \in P(M \setminus M_T) \mid c(p) < c(\hat{p})\}
	\qquad \text{and} \qquad
	\hat{B} = \{b \in B(A \setminus A_T) \mid v(b) > v(\hat{b})\}
	\enspace.
\]
Intuitively, $\hat{P}$ is the set of the assignable users, and $\hat{B}$ is the set of the assignable slots. It is important to note that $\hat{P}$ and $\hat{B}$ are both empty whenever $\hat{p}$ and $\hat{b}$ are dummy user and slot, respectively. We also define two additional sets $A_L$ and $M_L$ as follows. Let $f$ be a random variable distributed according to the binomial distribution $\cB(|A \setminus A_T| + |M \setminus M_T|, \min\{16r^{-1} \cdot \sqrt[3]{\alpha}, 1\})$, and let $L$ be the set of the last $f$ entities in $\sigma_E$ (or equivalently, the last $f$ entities to arrive). The sets $A_L$ and $M_L$ are then defined as $A_L = A \cap L$ and $M_L = M \cap L$.

\begin{lemma} \label{lem:event_summary}
There exists an event $\cE$ of probability at least $1 - 10e^{-2/\sqrt[3]{\alpha}}$ such that $\cE$ implies the following:
\begin{center}
\begin{tabular}{llll}
	(i)  & $\tilde{B} \setminus B(A_T) \subseteq \hat{B}$ \hspace{3cm} & (iii) & $|\hat{P} \setminus P(M_L)| \leq |\hat{B}|$\\
	(ii) & $\tilde{P} \setminus P(M_T) \subseteq \hat{P}$ \hspace{3cm} & (iv)  & $|\hat{B} \setminus B(A_L)| \leq |\hat{P}|$ \\
	(v)  & \multicolumn{3}{p{11cm}}{$c(p) \leq \ell(P, B) \leq v(b)$ for every user $p \in \hat{P}$ and slot $b \in \hat{B}$, where $\ell(P, B)$ is a value which is independent of the random coins of {\OPM} and obeys $c(p) \leq \ell(P, B) \leq v(b)$ for every $p \in P_o$ and $b \in B_o$.}
\end{tabular}
\end{center}
\end{lemma}

The proof of Lemma~\ref{lem:event_summary} is very similar to the proof of Lemma~4.6 in~\cite{FG16}, and thus, we defer it to Appendix~\ref{app:event_proof}. In the rest of this section we explain how the competitive ratio of {\OPM} follows from Lemma~\ref{lem:event_summary}. Let $\hat{S}$ be the assignment produced by {\OPM}.

\begin{lemma} \label{lem:joined_event}
The event $\cE$ implies the following inequality:
\[
	\GfT(\hat{S})
	\geq
	\sum_{\substack{b \in \tilde{B} \\ b \not \in B(A_T \cup A_L)}} \mspace{-18mu}[v(b) - \ell(P, B)] + \sum_{\substack{p \in \tilde{P} \\ p \not \in P(M_T \cup M_L)}} \mspace{-18mu}[\ell(P, B) - c(p)]
	\enspace.
\]
\end{lemma}
\begin{proof}
Lemma~\ref{lem:event_summary} shows that given $\cE$ we have $|\hat{P} \setminus P(M_L)| \leq |\hat{B}|$, hence, Observation~\ref{obs:invariant} implies that {\OPM} assigns at least $|\hat{P} \setminus P(M_L)|$ users. Additionally, since {\OPM} assigns users of mediators from $M_L$ only after all the assignable users of mediators from $M \setminus (M_T \cup M_L)$ are assigned to slots we get that all the users of $\hat{P} \setminus P(M_L)$ are assigned by $\hat{S}$ given $\cE$. On the other hand, Lemma~\ref{lem:event_summary} also shows that given $\cE$ all the users of $\tilde{P} \setminus P(M_T)$ belong to $\hat{P}$, and thus, the users of $\tilde{P} \setminus P(M_T \cup M_L)$ are all assigned by $\hat{S}$. A similar argument shows that the slots of $\tilde{B} \setminus B(A_T \cup A_L)$ are all assigned users by $\hat{S}$ given $\cE$. Finally, observe that $\cE$ also implies that $c(p) \leq \ell(P, B) \leq v(b)$ for every pair $(p, b) \in \hat{S} \subseteq \hat{P} \times \hat{B}$.

In the rest of the proof we assume that $\cE$ happens. Consider an ordered pair $(p, b) \in \hat{S}$. Then, the contribution of $(p, b)$ to $\GfT(\hat{S})$ is:
\[
	v(b) - c(p)
	=
	[v(b) - \ell(P, B)] + [\ell(P, B) - c(p)]
	\enspace.
\]
By the above discussion, the two terms that appear in brackets on the right hand side of the last equation are both positive. This allows us to lower bound the gain from trade of $\hat{S}$ as follows:
\begin{align*}
	\GfT(\hat{S})
	={} &
	\sum_{(p, b) \in \hat{S}} [v(b) - c(p)]
	=
	\sum_{(p, b) \in \hat{S}} \{[v(b) - \ell(P, B)] + [\ell(P, B) - c(p)]\}\\
	\geq{} &
	\sum_{\substack{b \in \tilde{B} \\ b \not \in B(A_T \cup A_L)}} \mspace{-18mu}[v(b) - \ell(P, B)] + \sum_{\substack{p \in \tilde{P} \\ p \not \in P(M_T \cup M_L)}} \mspace{-18mu}[\ell(P, B) - c(p)]
	\enspace.
	\qedhere
\end{align*}
\end{proof}

\begin{corollary} \label{cor:OPM_competitive_ratio}
{\OPM} is at least $(1 - r - 22r^{-1} \cdot \sqrt[3]{\alpha} - 10e^{-2/\sqrt[3]{\alpha}})$-competitive.
\end{corollary}
\begin{proof}
The corollary is trivial when $r + 22r^{-1} \cdot \sqrt[3]{\alpha} + 10e^{-2/\sqrt[3]{\alpha}} > 1$. Thus, we assume in this proof $r + 22r^{-1} \cdot \sqrt[3]{\alpha} + 10e^{-2/\sqrt[3]{\alpha}} \leq 1$. For every two sets $M' \subseteq M$ and $A' \subseteq A$ of mediators and advertisers, respectively, let $\Val(M', A')$ denote the expression:
\[
	\sum_{b \in \tilde{B} \setminus B(A')} [v(b) - \ell(P, B)] + \sum_{p \in \tilde{P} \setminus P(M')} [\ell(P, B) - c(p)]
	\enspace.
\]
The definition of $\ell(P, B)$ guarantees that $v(b) - \ell(P, B) \geq 0$ and $\ell(P, B) - c(p) \geq 0$ for every $b \in \tilde{B} \subseteq B_o$ and $p \in \tilde{P} \subseteq P_o$. Thus, $\Val(M', A') \leq \Val(\varnothing, \varnothing)$ for every two sets $M' \subseteq M$ and $A' \subseteq A$. Additionally, it is well-known that the way $t$ is chosen guarantees that every entity of $M \cup A$ belongs to $T$ with probability $r$, independently. A proof of this fact can be found, \eg, as Lemma~A.1 in~\cite{FSZ15}. Similarly, every entity of $M \cup A$ that does not belong to $T$ is added to $L$ with probability $\min\{1, 16r^{-1} \cdot \sqrt[3]{\alpha}\} = 16r^{-1} \cdot \sqrt[3]{\alpha}$, independently. Hence, every user (slot) of $\tilde{P}$ ($\tilde{B}$) belongs to $\tilde{P} \setminus P(M_T \cup M_L)$ ($\tilde{B} \setminus B(A_T \cup A_L)$) with probability
\[
	(1 - r)(1 - 16r^{-1} \cdot \sqrt[3]{\alpha})
	\geq
	1 - r - 16r^{-1} \cdot \sqrt[3]{\alpha}
	\enspace.
\]
Therefore,
\begin{align*}
	\bE[\Val(M_T \cup M_L, A_T \cup A_L)]
	\geq{} &
	(1 - r - 16r^{-1} \cdot \sqrt[3]{\alpha}) \cdot \sum_{b \in \tilde{B}} [v(b) - \ell(P, B)] \\ &+ (1 - r - 16r^{-1} \cdot \sqrt[3]{\alpha}) \cdot \sum_{p \in \tilde{P}} [\ell(P, B) - c(b)]\\
	={} &
	(1 - r - 16r^{-1} \cdot \sqrt[3]{\alpha}) \cdot \Val(\varnothing, \varnothing)
	\enspace.
\end{align*}
Using Lemma~\ref{lem:joined_event} and the observation that {\OPM} always produces assignments of non-negative gain from trade, we now get:
\begin{align} \label{eq:gain_from_trade_hat_S}
	\bE[\GfT(\hat{S})]
	={} &
	\Pr[\cE] \cdot \bE[\GfT(\hat{S}) \mid \cE] + \Pr[\neg \cE] \cdot \bE[\GfT(\hat{S}) \mid \neg \cE]\nonumber\\
	\geq{} &
	\Pr[\cE] \cdot \bE[\Val(M_T \cup M_L, A_T \cup A_L) \mid \cE]\nonumber\\
	={} &
	\bE[\Val(M_T \cup M_L, A_T \cup A_L)] - \Pr[\neg \cE] \cdot \bE[\Val(M_T \cup M_L, A_T \cup A_L) \mid \neg \cE]\nonumber\\
	\geq{} &
	(1 - r - 16r^{-1} \cdot \sqrt[3]{\alpha}) \cdot \Val(\varnothing, \varnothing) - \Pr[\neg \cE] \cdot \Val(\varnothing, \varnothing)\nonumber\\
	={} &
	[(1 - r - 16r^{-1} \cdot \sqrt[3]{\alpha}) - \Pr[\neg \cE]] \cdot \Val(\varnothing, \varnothing)
	\enspace.
\end{align}

Recall that $\Pr[\neg \cE] \leq 10e^{-2/\sqrt[3]{\alpha}}$ by Lemma~\ref{lem:event_summary}. Additionally, Observation~\ref{obs:tilde_elements} and the fact that $|\tilde{P}| = |\tilde{B}|$ by definition imply together:
\begin{align*}
	\Val(\varnothing, \varnothing)
	={} &
	\sum_{b \in \tilde{B}} [v(b) - \ell(P, B)] + \sum_{p \in \tilde{P}} [\ell(P, B) - c(p)]\\
	={} &
	\sum_{b \in \tilde{B}} v(b) - \sum_{p \in \tilde{P}} c(p)
	\geq
	(1 - 6r^{-1} \cdot \sqrt[3]{\alpha}) \cdot \GfT(S_c(P, A))
	\enspace.
\end{align*}
Plugging the last observations into Inequality~\eqref{eq:gain_from_trade_hat_S} gives:
\begin{align*}
	\bE[\GfT(\hat{S})]
	\geq{} &
	[(1 - r - 16r^{-1} \cdot \sqrt[3]{\alpha}) - \Pr[\neg \cE]] \cdot \Val(\varnothing, \varnothing)\\
	\geq{} &
	[(1 - r - 16r^{-1} \cdot \sqrt[3]{\alpha}) - 10e^{-2/\sqrt[3]{\alpha}}] \cdot (1 - 6r^{-1} \cdot \sqrt[3]{\alpha}) \cdot \GfT(S_c(P, A))\\
	\geq{} &
	(1 - r - 22r^{-1} \cdot \sqrt[3]{\alpha} - 10e^{-2/\sqrt[3]{\alpha}}) \cdot \GfT(S_c(P, B))
	\enspace.
\end{align*}
The corollary now follows by recalling that $S_c(P, B)$ is the assignment of users from $P$ to slots of $B$ which maximizes the gain from trade.
\end{proof}

\inArXiv{\bibliographystyle{plain}}
\inConference{\bibliographystyle{plainurl}}
\bibliography{../TwoSidedTrade}

\appendix
\makeatletter
\inConference{\edef\thetheorem{\expandafter\noexpand\thesection\@thmcountersep\@thmcounter{theorem}}}
\makeatother

\inConference{\section{Missing Proofs} \label{app:missing_proofs}

\begin{replemma}{lem:incentives_for_users}
For every user $p$, assuming the mediator $m$ of $p$ is truthful, {\OPM} is continuously individually rational for $p$, and truthfulness is a dominant strategy for her.
\end{replemma}
\incentivesForUsersProof

\begin{replemma}{lem:incentives_for_mediators}
For every mediator $m$, assuming the users of $m$ are truthful, {\OPM} is continuously individually rational for $m$, and truthfulness is a dominant strategy for him.
\end{replemma}
\incentivesForMediatorsProof

\begin{repobservation}{obs:tilde_elements}
$\sum_{b \in \tilde{B}} v(b) - \sum_{p \in \tilde{P}} c(p) \geq (1 - 6r^{-1} \cdot \sqrt[3]{\alpha}) \cdot \GfT(S_c(P, B))$.
\end{repobservation}
\tildeElementsProof}
\section{Proof of Lemma~\ref{lem:event_summary}} \label{app:event_proof}

In this section we prove Lemma~\ref{lem:event_summary}. Let us begin the proof with the following technical lemma (this lemma is identical to Lemma~4.9 in~\cite{FG16}. We repeat its proof here for completeness).

\begin{lemma} \label{lem:length_concentration}
Given a subset $B' \subseteq B_o$ and a probability $q \in [0, 1]$, let $B'[q]$ be a random subset of $B'$ constructed as follows: for every advertiser $a \in A$, independently, with probability $q$ the slots of advertiser $a$ that belong to $B'$ appear also in $B'[q]$. Then, for every $\beta \in (0, 1]$:
\[
	\Pr[||B'[q]| - q \cdot |B'|| \geq \beta \tau]
	\leq
	2e^{-2\beta^2/\alpha}
	\enspace.
\]
Similarly, given a subset $P' \subseteq P_o$ and a probability $q \in [0, 1]$, let $P'[q]$ be a random subset of $P'$ constructed as follows: for every mediator $m \in M$, independently, with probability $q$ the users of mediator $m$ that belong to $P'$ appear also in $P'[q]$. Then, for every $\beta \in (0, 1]$:
\[
	\Pr[||P'[q]| - q \cdot |P'|| \geq \beta \tau]
	\leq
	2e^{-2\beta^2/\alpha}
	\enspace.
\]
\end{lemma}
\begin{proof}
We prove the first inequality; the second inequality is analogous. First, observe that the lemma is trivial when $B' = \varnothing$ since $B' = \varnothing$ implies $||B'[q]| - q \cdot |B'|| = 0 < \beta \tau$. Thus, we may assume in the rest of the proof $B' \neq \varnothing$. For every advertiser $a \in A$, let $X_a$ be an indicator for the event that slots of $a$ appear in $B'[q]$. Then:
\[
	|B'[q]|
	=
	\sum_{a \in A} X_a \cdot |B' \cap B(a)|
	\enspace.
\]
The definition of $\alpha$ implies $|B(a)| \leq \alpha \tau$ for every advertiser $a \in A$, and thus, $0 \leq |B' \cap B(a)| \leq \alpha \tau$. Hence, by Hoeffding's inequality:
\begin{align*}
	\Pr[||B'[q]| - q \cdot |B'|| \geq \beta \tau]
	={} &
	\Pr[||B'[q]| - \bE[|B'[q]|]| \geq \beta \tau]
	\leq
	2e^{-\frac{2(\beta \tau)^2}{\sum_{a \in A} |B' \cap B(a)|^2}}\\
	\leq{} &
	2e^{-\frac{2(\beta \tau)^2}{\alpha \tau \cdot \sum_{a \in A} |B' \cap B(a)|}}
	=
	2e^{-\frac{2\beta^2 \tau}{\alpha \cdot |B'|}}
	\leq
	2e^{-\frac{2\beta^2 \tau}{\alpha \cdot |B_o|}}
	=
	2e^{-\frac{2\beta^2}{\alpha}}
	\enspace.
	\qedhere
\end{align*}
\end{proof}

\hypertargettop{event:E_prime}Let $\cE'$ be the event that the following inequalities are all true (at the same time):
\begin{center}
\begin{tabular}{llll}
	(i)  & $||B_o \cap B(A_T)| - r \cdot |B_o|| \leq \sqrt[3]{\alpha} \cdot \tau$ \hspace{1cm} & (iii) & $||\tilde{B} \cap B(A_T)| - r \cdot |\tilde{B}|| \leq \sqrt[3]{\alpha} \cdot \tau$\\
	(ii) & $||P_o \cap P(M_T)| - r \cdot |P_o|| \leq \sqrt[3]{\alpha} \cdot \tau$              & (iv)  & $||\tilde{P} \cap P(M_T)| - r \cdot |\tilde{P}|| \leq \sqrt[3]{\alpha} \cdot \tau$
\end{tabular}
\end{center}

\begin{observation} \label{obs:E_prime_probability}
$\Pr[\cE'] \geq 1 - 8e^{-2/\sqrt[3]{\alpha}}$.
\end{observation}
\begin{proof}
As explained in the proof of Corollary~\ref{cor:OPM_competitive_ratio}, $T$ contains every entity of $M \cup A$ with probability $r$, independently. This means that $B_o \cap B(A_T)$, $\tilde{B} \cap B(A_T)$, $P_o \cap P(M_T)$ and $\tilde{P} \cap P(M_T)$ have the same distributions as $B_o[r]$, $\tilde{B}[r]$, $P_o[r]$ and $\tilde{P}[r]$, respectively. Moreover, by definition, $\tilde{B} \subseteq B_o$ and $\tilde{P} \subseteq P_o$. Hence, by Lemma~\ref{lem:length_concentration}, each one of the four inequalities defining $\cE'$ holds with probability at least $	1 - 2e^{-2/\sqrt[3]{\alpha}}$. The observation now follows by the union bound.
\end{proof}

Next, we need the following useful observation (this observation is analogous to Observation~4.11 in~\cite{FG16}, and both observations share identical proofs. We repeat the proof here for completeness).

\begin{observation} \label{obs:length_characterization}
It always holds that:
\begin{align*}
	\min\{|P_o \cap P(M_T)|, |B_o \cap B(A_T)|\}
	\leq\inConference{{} &}
	|S_c(P(M_T), B(A_T))|\inConference{\\}
	\leq\inConference{{} &}
	\max \{|P_o \cap P(M_T)|, |B_o \cap B(A_T)|\}
	\enspace.
\end{align*}
\end{observation}
\begin{proof}
Let $p_\tau$ and $b_\tau$ be the user and slot at location $\tau$ of $S_c(P, B)$, respectively. The definition of a canonical assignment guarantees that we have $c(p_{\tau}) < v(b_{\tau})$. Additionally, the slots of $B_o$ all appear in locations $1$ to $\tau$ of $S_c(P, B)$, and thus, they all have values at least as large as $v(b_\tau)$. Similarly, the users of $P_o$ all have costs at most as large as $c(p_\tau)$. Combining these observations, we get: $c(p) \leq c(p_\tau) < v(b_\tau) \leq v(b)$ for every $p \in P_o$ and $b \in B_o$.

The slots at locations $1$ to $|B_o \cap B(A_T)|$ of $S_c(P(M_T), B(A_T))$ all belong to $B_o$ since $B_o$ contains the $\tau$ slots with the largest values. Similarly, the users at locations $1$ to $|P_o \cap P(M_T)|$ belong to $P_o$. Combining both observations, we get that for every location $1 \leq i \leq \min\{|P_o \cap P(M_T)|, |B_o \cap B(A_T)|\}$, the user $p'_i$ at location $i$ of $S_c(P(M_T), B(A_T))$ and the slot $b'_i$ at this location belong to $P_o$ and $B_o$, respectively, and thus, $c(p'_i) < v(b'_i)$. Hence, by the definition of a canonical assignment, the pair $(p'_i, b'_i)$ belongs to $S_c(P(M_T), B(A_T))$ for every $1 \leq i \leq \min\{|P_o \cap P(M_T)|, |B_o \cap B(A_T)|\}$; which completes the proof of the first inequality we need to prove.

Assume towards a contradiction that the second inequality we need to prove is wrong. In other words, we assume $|S_c(P(M_T), B(A_T))| > \max \{|P_o \cap P(M_T)|, |B_o \cap B(A_T)|\}$. Let $j = \max \{|P_o \cap P(M_T)|, |B_o \cap B(A_T)|\} + 1$, and let $p'_j$ and $b'_j$ be the user and slot at location $j$ of $S_c(P(M_T), B(A_T))$, respectively. Our assumption implies that $(p'_j , b'_j)$ belongs to $S_c(P(M_T), B(A_T))$, and thus, $c(p'_j) < v(b'_j)$. On the other hand, only the users at locations $1$ to $|P_o \cap P(M_T)|$ of $S_c(P(M_T), B(A_T))$ belong to $P_o$, hence, $p'_j$ does not belong to $P_o$. The user with the lowest cost that does not belong to $P_o$ is the user $p_{\tau + 1}$ at location $\tau + 1$ of $S_c(P, B)$. Thus, we get: $c(p'_j) \geq c(p_{\tau + 1})$. Similarly, we can also get $v(b'_j) \leq v(b_{\tau + 1})$, where $b_{\tau + 1}$ is the slot at location $\tau + 1$ of $S_c(P, B)$. Combining the above inequalities gives:
\[
	c(p_{\tau + 1})
	\leq
	c(p'_j)
	<
	v(p'_j)
	\leq
	v(b_{\tau + 1})
	\enspace,
\]
which contradicts the fact that $p_{\tau + 1}$ is not assigned to $b_{\tau + 1}$ by the canonical assignment $S_c(P, B)$.
\end{proof}

The next few claims use the last observation to prove a few properties that hold given $\cE'$. 

\begin{lemma} \label{lem:upper_inclusion}
The event \hyperlink{event:E_prime}{$\cE'$} implies: $\hat{P} \subseteq P_o$ and $\hat{B} \subseteq B_o$.
\end{lemma}
\begin{proof}
We prove the first inclusion. The other inclusion is analogous. Observation~\ref{obs:length_characterization} and the definition of $\cE'$ imply:
\begin{align*}
	|S_c(P(M_T), B(A_T))|
	\leq{} &
	\max \{|P_o \cap P(M_T)|, |B_o \cap B(A_T)|\}\\
	\leq{} &
	\max\{r \cdot |P_o| + \sqrt[3]{\alpha} \cdot \tau, r \cdot |B_o|  + \sqrt[3]{\alpha} \cdot \tau\}
	=
	(r + \sqrt[3]{\alpha})\tau
	\enspace.
\end{align*}
Using the definition of $\cE'$ again gives:
\begin{align*}
	|P_o \cap P(M_T)|
	\geq{} &
	r \cdot |P_o| - \sqrt[3]{\alpha} \cdot \tau
	=
	(r - \sqrt[3]{\alpha})\tau\\
	\geq{} &
	(1 - 2r^{-1} \cdot \sqrt[3]{\alpha}) \cdot (r + \sqrt[3]{\alpha})\tau
	\geq
	(1 - 2r^{-1} \cdot \sqrt[3]{\alpha}) \cdot |S_c(P(M_T), B(A_T))|
	\enspace,
\end{align*}
which implies, since $|P_o \cap P(M_T)|$ is integral,
\begin{equation} \label{eq:threshold_in_original}
	|P_o \cap P(M_T)|
	\geq
	\lceil (1 - 2r^{-1} \cdot \sqrt[3]{\alpha}) \cdot |S_c(P(M_T), B(A_T))| \rceil
	\enspace.
\end{equation}

If $\hat{p}$ is a dummy user then $\hat{P}$ is empty, which makes the claim $\hat{P} \subseteq P_o$ trivial. Thus, we may assume that $\hat{p}$ is the user at location $\lceil (1 - 2r^{-1} \cdot \sqrt[3]{\alpha}) \cdot |S_c(P(M_T), B(A_T))| \rceil$ of the canonical assignment $S_c(P(M_T), B(A_T))$. Hence, Inequality~\eqref{eq:threshold_in_original} and the observation that the users of $P_o \cap P(M_T)$ are the users with the lowest costs in $P(M_T)$ imply together that $\hat{p}$ belongs to the set $P_o \cap P(M_T) \subseteq P_o$. 
On the other hand, $P_o$ contains the $\tau$ users with the lowest costs. Hence, every user with a cost lower than $c(\hat{p})$ must be in $P_o$ since $\hat{p}$ is in $P_o$. The lemma now follows by observing that the definition of $\hat{P}$ implies $c(p) < c(\hat{p})$ for every user $p \in \hat{P}$. 
\end{proof}

\begin{corollary}\label{cor:middle_value}
There exists a value $\ell(P, B)$ independent of the random coins of {\OPM} such that:
\begin{\itemizeEnvironment}
	\item $c(p) \leq \ell(P, B) \leq v(b)$ for every user $p \in P_o$ and slot $b \in B_o$
	\item Whenever the event \hyperlink{event:E_prime}{$\cE'$} occurs, $c(p) \leq \ell(P, B) \leq v(b)$ for every user $p \in \hat{P}$ and slot $b \in \hat{B}$.
\end{\itemizeEnvironment}
\end{corollary}
\begin{proof}
Let $\ell(P, B)$ be the value of the slot at location $\tau$ of the canonical assignment $S_c(P, B)$. Clearly, $\ell(P, B)$ is independent of the random coins of {\OPM}, as required. Additionally, for every slot $b \in B_o$ it holds that $v(b) \geq \ell(P, B)$ since $b$ must be located at some location of $S_c(P, B)$ between $1$ and $\tau$. On the other hand, let $p_\tau$ be the user at location $\tau$ of $S_c(P, B)$. Since the size of $S_c(P, B)$ is $\tau$, $p_\tau$ must be assigned to the slot at location $\tau$ of $S_c(P, B)$, which implies $c(p_{\tau}) \leq \ell(P, B)$.\footnote{In fact, we even have $c(p_{\tau}) < \ell(P, B)$ since the tie-breaking rule defined in Section~\ref{sec:comparison} guarantees that the value of a slot is never equal to the cost of a user.} Moreover, for every user $p \in P_o$ it holds that $c(p) \leq c(p_\tau) \leq \ell(P, B)$ since $p$ must be located at some location of $S_c(P, B)$ between $1$ and $\tau$.

The corollary now follows since Lemma~\ref{lem:upper_inclusion} shows that the event $\cE'$ implies that every user $p \in \hat{P}$ belongs also to $P_o$, and every slot $b \in \hat{B}$ belongs also to $B_o$.
\end{proof}

\begin{lemma} \label{lem:lower_inclusion}
The event \hyperlink{event:E_prime}{$\cE'$} implies: $\tilde{P} \setminus P(M_T) \subseteq \hat{P}$ and $\tilde{B} \setminus B(A_T) \subseteq \hat{B}$.
\end{lemma}
\begin{proof}
We prove the first inclusion. The other inclusion is analogous. The claim about $\tilde{P} \cap P(M \setminus M_T)$ is trivial when $\tilde{P}$ is empty. Thus, we can assume throughout the proof that $\tilde{P}$ is non-empty. Observation~\ref{obs:length_characterization} and the definition of $\cE'$ imply:
\begin{align*}
	|S_c(P(M_T), B(A_T))|
	\geq{} &
	\min \{|P_o \cap P(M_T)|, |B_o \cap B(A_T)|\}\\
	\geq{} &
	\min\{r \cdot |P_o| - \sqrt[3]{\alpha} \cdot \tau, r \cdot |B_o| - \sqrt[3]{\alpha} \cdot \tau\}
	=
	(r - \sqrt[3]{\alpha})\tau
	\enspace.
\end{align*}
Recall that $\alpha \geq \tau^{-1}$, and thus, $\sqrt[3]{\alpha} \cdot \tau \geq 1$. Using this inequality and the definitions of $\cE'$ and $\tilde{P}$ now gives:
\begin{align*}
	|\tilde{P} \cap P(M_T)|
	\leq{} &
	r \cdot |\tilde{P}| + \sqrt[3]{\alpha} \cdot \tau
	=
	r \cdot \lceil (1 - 6r^{-1} \cdot \sqrt[3]{\alpha})\tau \rceil + \sqrt[3]{\alpha} \cdot \tau\inConference{\\}
	\leq\inConference{{} &}
	(r - 4\sqrt[3]{\alpha})\tau - 1 + \sqrt[3]{\alpha} \cdot \tau\inArXiv{\\}
	=\inArXiv{{} &}
	(r - 3\sqrt[3]{\alpha})\tau - 1\inConference{\\}
	\leq\inConference{{} &}
	(1 - 2r^{-1} \cdot \sqrt[3]{\alpha}) \cdot (r - \sqrt[3]{\alpha})\tau - 1\\
	\leq{} &
	(1 - 2r^{-1} \cdot \sqrt[3]{\alpha}) \cdot |S_c(P(M_T), B(A_T))| - 1
	\enspace,
\end{align*}
which implies, since $|\tilde{P} \cap P(M_T)|$ is integral,
\begin{align} \label{eq:threshold_above_tilde}
	|\tilde{P} \cap P(M_T)|
	\leq{} &
	\lceil (1 - 2r^{-1} \cdot \sqrt[3]{\alpha}) \cdot |S_c(P(M_T), B(A_T))| \rceil - 1\\
	<{} &
	\lceil (1 - 2r^{-1} \cdot \sqrt[3]{\alpha}) \cdot |S_c(P(M_T), B(A_T))| \rceil \nonumber
	\enspace.
\end{align}

Inequality~\eqref{eq:threshold_above_tilde} and the observation that the users of $\tilde{P} \cap P(M_T)$ are the users with the lowest costs in $P(M_T)$ imply together that $\hat{p}$ is a user of $P(M_T)$ which does not belong to $\tilde{P} \cap P(M_T)$, and therefore, does not belong to $\tilde{P}$ either. 
On the other hand, $\tilde{P}$ contains the $\lceil (1 - 6r^{-1} \cdot \sqrt[3]{\alpha})\tau \rceil$ users with the lowest costs. Hence, every user $p \in \tilde{P}$ has a cost smaller than $c(\hat{p})$ since $\hat{p}$ does not belong to $\tilde{P}$. The lemma now follows by observing that the definition of $\hat{P}$ implies that $p \in \hat{P}$ for every user $p \in P(M \setminus M_T)$ obeying $c(p) < c(\hat{p})$. 
\end{proof}

\begin{corollary} \label{cor:size_bounds}
The event \hyperlink{event:E_prime}{$\cE'$} implies: $-7r^{-1} \cdot \sqrt[3]{\alpha} \cdot \tau \leq |\hat{P}| - (1 - r)\tau \leq \sqrt[3]{\alpha} \cdot \tau$ and $-7r^{-1} \cdot \sqrt[3]{\alpha} \cdot \tau \leq |\hat{B}| - (1 - r)\tau \leq \sqrt[3]{\alpha} \cdot \tau$.
\end{corollary}
\begin{proof}
We prove here only the bounds on the size of $\hat{P}$. The bounds on the size of $\hat{B}$ are analogous. By Lemma~\ref{lem:upper_inclusion}, $\hat{P} \subseteq P_o$. On the other hand, by definition, $\hat{P} \subseteq P(M \setminus M_T)$. Thus, we get: $\hat{P} \subseteq P_o \setminus P(M_T)$. Combining this inclusion with the definition of $\cE'$ gives:
\begin{align*}
	|\hat{P}|
	\leq{} &
	|P_o \setminus P(M_T)|
	=
	|P_o| - |P_o \cap P(M_T)|\\
	\leq{} &
	|P_o| - [r \cdot |P_o| - \sqrt[3]{\alpha} \cdot \tau]
	=
	(1 - r) \cdot |P_o| + \sqrt[3]{\alpha} \cdot \tau
	=
	(1 - r)\tau + \sqrt[3]{\alpha} \cdot \tau
	\enspace.
\end{align*}
On the other hand, by Lemma~\ref{lem:lower_inclusion} and the definition of $\cE'$,
\begin{align*}
	|\hat{P}|
	\geq{} &
	|\tilde{P} \setminus P(M_T)|
	=
	|\tilde{P}| - |\tilde{P} \cap P(M_T)|
	\geq
	|\tilde{P}| - [r \cdot |\tilde{P}| + \sqrt[3]{\alpha} \cdot \tau]\\
	={} &
	(1 - r) \cdot |\tilde{P}| - \sqrt[3]{\alpha} \cdot \tau
	\geq
	(1 - r) \cdot \lceil (1 - 6r^{-1} \cdot \sqrt[3]{\alpha})\tau \rceil - \sqrt[3]{\alpha} \cdot \tau\inConference{\\}
	\geq\inConference{{} &}
	(1 - r)\tau - 7r^{-1} \cdot \sqrt[3]{\alpha} \cdot \tau
	\enspace.
	\qedhere
\end{align*}
\end{proof}

We can now define the event $\cE$ referred to by Lemma~\ref{lem:event_summary}. The event $\cE$ is the event that $\cE'$ happens and in addition the following two inequalities also hold:
\begin{center}
\begin{tabular}{llll}
	(i)  & $|\hat{B} \setminus B(A_L)| \leq |\hat{P}|$ \hspace{1cm} & (ii) & $|\hat{P} \setminus P(M_L)| \leq |\hat{B}|$
\end{tabular}
\end{center}

We are now ready to prove Lemma~\ref{lem:event_summary}. For ease of the reading, we first repeat the lemma itself.

\begin{replemma}{lem:event_summary}
There exists an event $\cE$ of probability at least $1 - 10e^{-2/\sqrt[3]{\alpha}}$ such that $\cE$ implies the following:
\begin{center}
\begin{tabular}{llll}
	(i)  & $\tilde{B} \setminus B(A_T) \subseteq \hat{B}$ \hspace{3cm} & (iii) & $|\hat{P} \setminus P(M_L)| \leq |\hat{B}|$\\
	(ii) & $\tilde{P} \setminus P(M_T) \subseteq \hat{P}$ \hspace{3cm} & (iv)  & $|\hat{B} \setminus B(A_L)| \leq |\hat{P}|$ \\
	(v)  & \multicolumn{3}{p{11cm}}{$c(p) \leq \ell(P, B) \leq v(b)$ for every user $p \in \hat{P}$ and slot $b \in \hat{B}$, where $\ell(P, B)$ is a value which is independent of the random coins of {\OPM} and obeys $c(p) \leq \ell(P, B) \leq v(b)$ for every $p \in P_o$ and $b \in B_o$.}
\end{tabular}
\end{center}
\end{replemma}
\begin{proof}
By definition, the event $\cE$ implies the inequalities: $|\hat{B} \setminus B(A_L)| \leq |\hat{P}|$ and $|\hat{P} \setminus P(M_L)| \leq |\hat{B}|$. Additionally, $\cE$ implies the event $\cE'$, which, by Corollary~\ref{cor:middle_value} and Lemma~\ref{lem:lower_inclusion}, implies the other things that should follow from $\cE$ by the lemma. Hence, the only thing left to prove is that the probability of $\cE$ is at least $1 - 10e^{-2/\sqrt[3]{\alpha}}$.

If $16r^{-1} \cdot \sqrt[3]{\alpha} \geq 1$, then $L$ contains all the entities arriving after the observation phase, which implies $A_L = A \setminus A_T$ and $M_L = M \setminus M_T$; and thus, the two inequalities $|\hat{B} \setminus B(A_L)| \leq |\hat{P}|$ and $|\hat{P} \setminus P(M_L)| \leq |\hat{B}|$ are trivial in this case. Hence, the events $\cE$ and $\cE'$ are equivalent when $16r^{-1} \cdot \sqrt[3]{\alpha} \geq 1$, and therefore, the probability of $\cE$ is at least $1 - 8e^{-2/\sqrt[3]{\alpha}}$ by Observation~\ref{obs:E_prime_probability}. Thus, it is safe to assume in the rest of the proof that $16r^{-1} \cdot \sqrt[3]{\alpha} < 1$.

Our plan is to prove the inequality $\Pr[\cE \mid \cE'] \geq 1 - 2e^{-2/\sqrt[3]{\alpha}}$. Notice that this inequality indeed implies the lemma since it implies:
\[
	\Pr[\cE]
	=
	\Pr[\cE'] \cdot \Pr[\cE \mid \cE']
	\geq
	(1 - 8e^{-2/\sqrt[3]{\alpha}}) \cdot (1 - 2e^{-2/\sqrt[3]{\alpha}})
	\geq
	1 - 10e^{-2/\sqrt[3]{\alpha}}
	\enspace.
\]

The event $\cE'$ is fully determined by the sets $M_T$ and $A_T$. Thus, it is enough to show that for every fixed choice of these sets for which the event $\cE'$ holds, the event $\cE$ holds with probability at least $1 - 2e^{-2/\sqrt[3]{\alpha}}$. Notice that the sets $\hat{P}$ and $\hat{B}$ become deterministic once we fix the choice of $M_T$ and $A_T$. Hence, either $|\hat{B}| \leq |\hat{P}|$, which implies that the inequality $|\hat{B} \setminus B(A_L)| \leq |\hat{P}|$ holds regardless of the choice of $A_L$, or $|\hat{P}| \leq |\hat{B}|$, which implies that the inequality $|\hat{P} \setminus P(M_L)| \leq |\hat{B}|$ holds regardless of the choice of $M_L$. In both cases, all we need to show is that the other inequality holds with probability at least $1 - 2e^{-2/\sqrt[3]{\alpha}}$ over the random choice of $A_L$ and $M_L$.

Let us assume, without loss of generality, that $|\hat{B}| \leq |\hat{P}|$. By the above discussion, all we need to prove is that $\Pr[|\hat{P} \setminus P(M_L)| \leq |\hat{B}|] \geq 1 - 2e^{-2/\sqrt[3]{\alpha}}$, where the probability is over the random choice of $M_L$. By Corollary~\ref{cor:size_bounds}:
\begin{align*}
	\Pr[|\hat{P} \setminus P(M_L)| > |\hat{B}|]\inConference{&{}}
	\leq\inArXiv{{} &}
	\Pr[|\hat{P} \setminus P(M_L)| > (1 - r)\tau - 7r^{-1} \cdot \sqrt[3]{\alpha} \cdot \tau]\\
	\leq{} &
	\Pr[|\hat{P} \setminus P(M_L)| > (1 - 16r^{-1} \cdot \sqrt[3]{\alpha})((1 - r)\tau + \sqrt[3]{\alpha} \cdot \tau) + \sqrt[3]{\alpha} \cdot \tau]\\
	\leq{} &
	\Pr[|\hat{P} \setminus P(M_L)| > (1 - 16r^{-1} \cdot \sqrt[3]{\alpha}) \cdot |\hat{P}| + \sqrt[3]{\alpha} \cdot \tau]
	\enspace.
\end{align*}
As explained in the proof of Corollary~\ref{cor:OPM_competitive_ratio}, $M_L$ contains every mediator of $M \setminus M_T$ with probability $16r^{-1} \cdot \sqrt[3]{\alpha}$, independently. Hence, $\hat{P} \setminus P(M_L)$ has the same distribution as $\hat{P}[1 - 16r^{-1} \cdot \sqrt[3]{\alpha}]$. Therefore, by Lemma~\ref{lem:length_concentration}:
\begin{align*}
	\Pr[|\hat{P} \setminus P(M_L)| > |\hat{B}|]
	\leq{} &
	\Pr[|\hat{P} \setminus P(M_L)| > (1 - 16r^{-1} \cdot \sqrt[3]{\alpha}) \cdot |\hat{P}| + \sqrt[3]{\alpha} \cdot \tau]\\
	\leq{} &
	\Pr[||\hat{P}(1 - 16r^{-1} \cdot \sqrt[3]{\alpha})| - (1 - 16r^{-1} \cdot \sqrt[3]{\alpha}) \cdot |\hat{P}|| > \sqrt[3]{\alpha} \cdot \tau]\\
	\leq{} &
	2e^{-2/\sqrt[3]{\alpha}}
	\enspace.
	\qedhere
\end{align*}
\end{proof}

\end{document}